\documentclass[prd,nofootinbib,showpacs,twocolumn,floatfix]{revtex4}
\usepackage{latexsym}
\usepackage{dcolumn}
\RequirePackage{graphicx}

\newcommand{\MET}{{\slash\!\!\!\!E_T}}
\newcommand{\DZero}{D$\slash\!\!\!0$}
\newcommand{\fbi}{\mathrm{fb}^{-1}}

\newcommand{\be}{\begin{equation}}
\newcommand{\ee}{\end{equation}}
\newcommand{\bea}{\begin{eqnarray}}
\newcommand{\eea}{\end{eqnarray}}

\begin{document}

\preprint{IIT-CAPP-11-02}

\title{A standard model explanation of a CDF dijet excess in $Wjj$}

\author{Zack~Sullivan}
\email{Zack.Sullivan@IIT.edu}
\affiliation{Illinois Institute of Technology,
Chicago, Illinois 60616-3793, USA}
\author{Arjun~Menon}
\affiliation{Illinois Institute of Technology,
Chicago, Illinois 60616-3793, USA}

\date{\today}

\begin{abstract}
  We demonstrate the recent observation of a peak in the dijet
  invariant mass of the $Wjj$ signal observed by the CDF Collaboration
  can be explained as the same upward fluctuation observed by CDF in
  single-top-quark production.  In general, both $t$-channel and
  $s$-channel single-top-quark production produce kinematically
  induced peaks in the dijet spectrum.  Since CDF used a Monte Carlo
  simulation to subtract the single-top backgrounds instead of data, a
  peak in the dijet spectrum is expected.  The \DZero\ Collaboration
  has a small upward fluctuation in their published $t$-channel data;
  and hence we predict they would see at most a small peak in the
  dijet invariant mass spectrum of $Wjj$ if they follow the same
  procedure as CDF.
\end{abstract}


\maketitle

The recent observation by CDF of a large peak in the
background-subtracted dijet spectrum of $Wjj$ has caused significant
excitement about a possible observation of beyond the standard model
physics \cite{Aaltonen:2011mk}.  The key question is whether this
excess is an artifact of the procedure used to subtract the standard
model background.  There are large excesses in a few exclusive
channels of the CDF single-top-quark data set \cite{Aaltonen:2010jr}
that can explain most, or possibly all, of the peak in the $Wjj$ data.
Hence there are two possibilities: there is evidence of unexpected
signals in multiple CDF measurements, or there is an upward
fluctuation in single-top-quark production with respect to Monte Carlo
that simply appears to be new physics.  Given the good agreement
between the standard model prediction and the \DZero\ data set for
single-top-quark production \cite{Abazov:2008kt}, we focus on the
latter case.

We first point out that events involving top-quark production
generically produce a peak in a dijet spectrum.  In the top-quark rest
frame, the $b$ in $t\to bW$ has an energy of $\sim 70$ GeV.  Hence,
there is a peak just below that energy in the transverse energy $E_T$
spectrum.  When combined with any other jet, where a jet is defined as
having $E_T$ greater than some threshold, the dijet invariant mass has
a peak above $E_{Tb}+E_{Tj}$.  This means that any analysis that finds
a peak in the dijet spectrum in the 100--160 GeV range will be
sensitive to how well the top-quark backgrounds are removed.  The CDF
analysis claims to have normalized the $t\bar t$ background to data,
but removed the single-top-quark backgrounds via Monte Carlo.  Thus,
we focus on single-top-quark production within the CDF data set.

In order to convert the measurement of single-top-quark production
into a prediction for $Wjj$, we start with the CDF single-top-quark
analysis \cite{Aaltonen:2010jr}.  This analysis was performed with the
same lepton trigger and a 3.2 $\fbi$ subset of the 4.3 $\fbi$ data
used for the $Wjj$ analysis.  We fit the cross sections observed by
CDF in four channels: $Wbj$, $Wbjj$, $Wbb$, and $Wbbj$ to a $K$-factor
times the next-to-leading order predictions from Ref.\
\cite{Sullivan:2004ie} for the exclusive final states.  In order to
account for cross-contamination in the 1-tag and 2-tag samples we
assume a 50\% average $b$-tagging efficiency.  In Table \ref{tab:fits}
we show the extracted $K$-factors with experimental errors.
Additional theoretical uncertainties are discussed below.

\begin{table}[htb]
\caption{$K$-factors for $t$-channel and $s$-channel single-top-quark 
exclusive final states extracted from fits to CDF data
\protect{\cite{Aaltonen:2010jr}} and the NLO predictions of
Ref.\ \protect{\cite{Sullivan:2004ie}}.
Uncertainties correspond to $1\sigma$ fluctuations in the CDF analysis.
\label{tab:fits}}
\begin{ruledtabular}
\begin{tabular}{ccccc}
Process & $Wbj$ & $Wbb$ & $Wbjj$ & $Wbbj$ \\
\hline
$t$-chan. & $0.6^{+0.3}_{-0.2}$ & $0.4^{+0.2}_{-0.2}$ & $0.9^{+0.8}_{-0.7}$ &
$2.0^{+1.5}_{-1.3}$ \\
$s$-chan. & $0.5^{+0.2}_{-0.1}$ & $3.8^{+2.1}_{-1.7}$ & $0.6^{+0.5}_{-0.4}$ &
$2.7^{+2.1}_{-1.8}$
\end{tabular}
\end{ruledtabular}
\end{table}

\begin{figure}[htb]
\centering
\includegraphics[width=0.9\columnwidth]{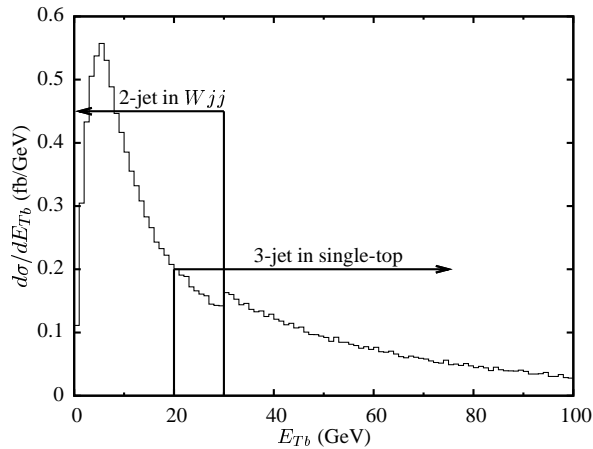}
\caption{Transverse energy spectrum of the additional $b$-jet in
$t$-channel single-top-quark production.  A 3-jet excess in single-top
(where $E_{Tj_3}>20$ GeV) is converted to a 2-jet excess in $Wjj$
(where $E_{Tj_3}<30$).
\label{fig:bjet}}
\end{figure}

It is interesting to note, the CDF extraction of the $t$-channel cross
section has a $K$-factor of 0.5 when averaging the $Wbj$ and $Wbb$
samples.  Hence, one might expect that subtracting the Monte Carlo
prediction from the data would lead to a deficit rather than an excess
in the dijet data.  However, the $Wbbj$ contribution to the
$t$-channel cross section has an excess of a factor of 2--3.5.  In the
current $Wjj$ exclusive final state, the cuts on the jets are raised
from $E_{Tj}>20$ GeV to $E_{Tj}>30$ GeV.  As mentioned in Ref.\
\cite{Sullivan:2004ie}, this converts a significant fraction of 3-jet
events from the single-top-quark sample to 2-jet events.  This is
demonstrated in Fig.\ \ref{fig:bjet}, where the area above 20 GeV is
similar to the area below 30 GeV in the distribution of the
initial-state radiated $b$-jet in $t$-channel production, which is
typically the third jet in the event.  It is critical to remember that
the distribution of that $b$-jet is only known to leading order (LO),
and is afflicted by the same large logarithms that enhance the
$t$-channel cross section.  Hence, when a sharp cut is applied to the
data, the uncertainty in the normalization of the exclusive 2-jet
state is nearly a factor of 2.

In addition to the measured excess, there are possible enhancements to
the exclusive 2-jet cross section due to other cuts on the data.  One
very sensitive cut is on the missing transverse energy $\MET > 25$
GeV.  Lowering the $\MET$ threshold in the theoretical prediction by 5
GeV changes the predicted single-top acceptance by $\sim 50\%$.  A 5
GeV shift is easily accommodated by the $\MET$ resolution, and the
fact that there are 1--2 $b$ hadrons in the final state.  These $b$'s
decay to neutrinos that add to the $\MET$.  In addition, the
initial-state radiation jets in single-top are frequently in less-well
instrumented regions of the detector and contribute to $\MET$.  All of
these effects enter in the same direction to enhance the predicted
cross section.

One visually striking feature of the CDF excess is the apparent peak
between 120--160 GeV.  If not for the low bin around 112 GeV, the
excess would go down to 96 GeV.  A direct scaling of CDF single-top
data would cover both this broader region and below.  However, the
position of the peak is very sensitive to the jet energy scale.  There
is a $25\%$ absolute correction times an additional $10\%$ out-of-cone
correction for the measured jet energies \cite{Bhatti:2005ai}.  These
large corrections lead to a 16\% systematic uncertainty in the jet
energy scale \cite{Aaltonen:2010jr}.  If some of the jets are $b$
jets, then the uncertainty may be larger.  The consequence is that the
predicted position of the dijet invariant-mass peak $M_{jj}$ can shift
higher toward the position of the CDF excess.

If the energy scale of the jets shifts upward, additional jets that
would have failed the dijet system transverse momentum cut $p_{T
  jj}>40$ GeV cut will now pass.  The theoretical prediction for the
$p_{T jj}$ spectrum is very sensitive to the cut chosen in the CDF
analysis.  It is important to remember that the $p_{T jj}$
distribution is only predicted at leading order.  As in Drell-Yan
production, resummation of initial-state radiation will slightly
harden this spectrum.  The dijet system will also recoil against any
third jet that passes the acceptance cut.  We estimate from our 3-jet
samples that these corrections can accommodate a 10--20 GeV upward
shift in the $p_{T jj}$ spectrum.  A 10 GeV shift would increase the
background acceptance by $\sim 50\%$.

One may wonder whether there is a large excess in the 2 $b$-tag CDF
dijet invariant mass.  CDF has measured that signal in an analysis to
search for Higgs production in $WH\to Wb\bar b$
\cite{Aaltonen:2009dh}.  There are two reasons we do not expect to see
a large excess in that study.  First, the deficit in $Wbb$ from
$t$-channel single-top is almost perfectly cancelled by the excess in
the $s$-channel single-top contribution.  The basic cuts in the Higgs
analysis are almost identical to the single-top-quark analysis, and so
there is no contamination from processes with additional jets.
Furthermore, in the CDF Higgs analysis, they normalize their
background subtraction to data.  Hence, any residual excess should be
removed.

\begin{figure}[htb]
\centering
\includegraphics[width=\columnwidth]{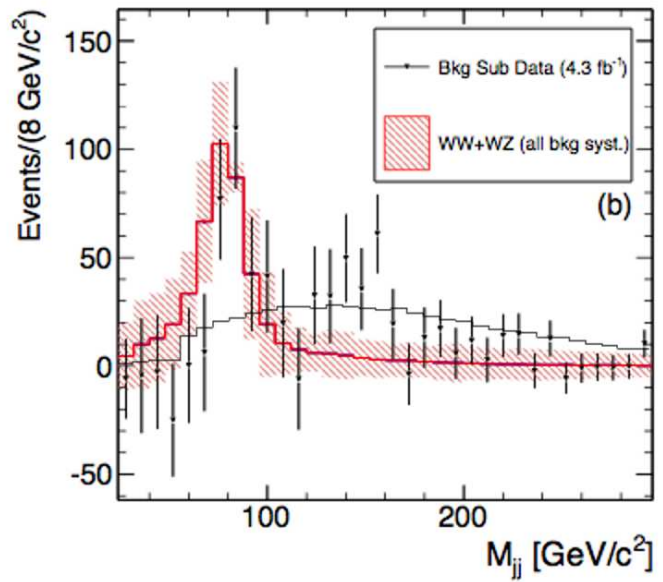}
\caption{Prediction of the single-top-quark excess (solid black line)
remaining in the dijet invariant mass $M_{jj}$ in $Wjj$ after
subtracting $1\times$ the standard model prediction from the CDF
single-top-quark data \protect{\cite{Aaltonen:2010jr}}.  $Wjj$ data
from Ref.\ \protect{\cite{Aaltonen:2011mk}}.
\label{fig:ourfit}}
\end{figure}

When we include the error in the \textit{prediction} of $\MET$ and jet
energy scale, we find that it is possible to reproduce the excess
distribution up to binning effects.  In Fig.\ \ref{fig:ourfit} we
superimpose on the CDF data a prediction obtained by boosting $\MET$
by 16\%, increasing the jet energy scale by 10\%, and mixing $3.5$
times the $t$-channel ($Wbbj$) background contribution to the 2-jet
sample, $-0.5$ times the $t$-channel ($Wbj$) contribution, and $5$
times the $s$-channel signal.  This combination corresponds to
slightly less than a $1\sigma$ fluctuation of the total single-top
background from CDF data minus 1 times the standard model theoretical
prediction.  Our net prediction (solid black histogram) for the
residual single-top-quark contribution to $M_{jj}$ reproduces the both
the excess between 120--160, and the excess on the right shoulder of
the $WW/WZ$ fit.  The modeling uncertainty in the normalization of
this curve is about a factor of 2.

In conclusion, we observe that the dijet invariant mass peak seen in
the recent CDF $Wjj$ cross section \cite{Aaltonen:2011mk} is
completely consistent with the excess observed in the CDF
single-top-quark analysis \cite{Aaltonen:2010jr}.  Both may be
explained by an upward fluctuation in the CDF data set of $s$-channel
single-top-quark production, and $t$-channel production accompanied by
an additional low-energy jet.  The latter process is poorly modeled by
Monte Carlo, and the apparent $t$-channel excess could simply be an
artifact of theoretical uncertainty.  Given the modest excess observed
by the \DZero\ Collaboration in their single-top-quark data set
\cite{Abazov:2008kt}, we predict the \DZero\ Collaboration would not
see a significant dijet invariant mass peak if they follow the CDF
procedure.

\begin{acknowledgments}
  This work is supported by the U.~S.\ Department of Energy under
  Contract No.\ DE-FG02-94ER40840.
\end{acknowledgments}

\end{document}